\documentclass{PoS}

\title{Beyond the Standard Model searches with top quarks at D0}

\ShortTitle{BSM searches with top quarks at D0}

\author{\speaker{Yvonne Peters for the D0 Collaboration}\thanks{Supported by the Royal Society.}\\
        University of Manchester\\
        Oxford Road   \\
        Manchester,  M13 9PL,  UK    \\
      E-mail: \email{peters@fnal.gov}}

\abstract{Due to its high mass and short lifetime, the top quark plays an important role in checking  the Standard Model of particle physics. In this report, we present a variety of searches for physics beyond the Standard Model,
  involving top quarks, at the D0 detector at the Fermilab Tevatron collider. Specifically,
  we present searches in top quark pair production, single top quark
  production and top quark decays. The search spectra discussed here involve a search for $t\bar{t}$
  resonances, associated production of Higgs bosons and $t\bar{t}$, charged
  Higgs bosons and heavy gauge  $W^{'}$ bosons. Furthermore, we  measure the
  forward-backward charge asymmetry and a ratio of  branching
  fractions. 

}

\FullConference{The 2009 Europhysics Conference on High Energy Physics,\\
		 July 16 - 22 2009\\
		 Krakow, Poland}

\begin{document}

\section{Introduction}
The heaviest known elementary particle today is the top quark, with a
mass of $173.1\pm1.3$~GeV~\cite{topmassworldaverage}. Due to its high
mass, the Yukawa coupling to the Higgs boson is expected to be
large. Furthermore, the top quark decays before it hadronizes, making
it a unique particle to study bare quarks. 

Since the discovery of the top quark in 1995 by CDF and
D0~\cite{CDF_obs, D0_obs}, several properties have
been investigated with increasing precision. For example, the $t\bar{t}$ production
cross section is now measured with a precision close to the
theoretical uncertainty~\cite{xseccombi}, and the measurement of the
top quark mass exceeds $1\%$ accuracy~\cite{topmassworldaverage}. Due to the high precision
measurements, high statistics and its particular properties the top
quark sector is an interesting sector to search for deviations from the Standard Model~(SM) expectation. 
Recently, the observation of
single top quarks was reported by CDF and D0~\cite{CDF_singletop,
  D0_singletop}, opening another channel to search for physics beyond
the SM. 

In the following, a selection of searches beyond the SM in $t\bar{t}$
production, top quark decay and single top quark production will be
presented. 

\section{\boldmath Searches for new physics in $t\bar{t}$ production}
At the Tevatron, top quark pair production occurs  85\% of the time via $q\bar{q}$
annihilation and with 15\% gluon-gluon fusion. Assuming SM production and
decay, the measured production cross section at a top mass of $170$~GeV is
$\sigma_{t\bar{t}}=8.18_{-0.87}^{+0.98}$~pb~\cite{xseccombi}.

For measurements of $t\bar{t}$ cross section and top properties the
$t\bar{t}$ final states are classified according to the decays of the
two $W$ bosons from the top and anti-top decay. We separate the final states into dileptonic,
semileptonic and allhadronic channels according to the number of leptons in the final state. If the lepton is a hadronic decaying
tau, the events are treated as separate channels ($\tau$+lepton and $\tau$+jets).

In the SM, no $t\bar{t}$ resonances exist, but many models like for
example Topcolor assisted technicolor models predict a
resonance.  Using the semileptonic final state, D0 performed a search for a narrow
 resonance $X$, manifesting as a  bump in the
invariant $t\bar{t}$ mass spectra~\cite{reso_d0}. With $3.6~{\rm fb}^{-1}$ of data, limits on
 $\sigma(p\bar{p}\rightarrow X)\times B(X\rightarrow t\bar{t})$ versus
   $M_{X}$ have been extracted. In the reference model of 
   Topcolor assisted technicolor a $Z^{'}$ can be excluded for  $M_{X}<820$~GeV. In Fig.~\ref{CPX_exclusion} the
 invariant mass distribution is shown, together with a $Z^{'}$ of mass
 650~GeV. 

Not only invariant mass spectra, but also the forward backward charge
asymmetry of $t\bar{t}$ events can give hints for deviations from the
SM. In next-to-leading-order (NLO) calculations, the asymmetry is
expected to be 5\%. With $1~{\rm fb}^{-1}$ of data, D0 performed a measurement of the asymmetry by
looking at the number of semileptonic events with a non-vanishing rapidity difference between
the top and the anti-top quark~\cite{asymmetry}. The
measured asymmetry of $A_{fb}=12\pm 8~{\rm (stat)}~\pm 1~{\rm (syst)}$\% is
consistent with the SM. 

Due to its high mass, the top quark is expected to have a large
coupling to the Higgs boson. An interesting search for the Higgs boson
is therefore to look for  associated production of $t \bar{t}$ and
Higgs. By studying events with a high number of jets, and especially a
high number of $b$-jets, D0 preformed a search for $t\bar{t}H$ with
$2.1~{\rm fb}^{-1}$ of data~\cite{ttH}. Compared to the SM prediction of
$\sigma_{t\bar{t}H}$, we reach an upper limit of 
$\sigma_{t\bar{t}H}/\sigma_{SM}< 48$ at 95\% C.L. assuming a Higgs mass of 105~GeV. In models
beyond the SM, the $t\bar{t}H$ cross section can be enhanced, for
example, if a $t^{ '}$ is produced via a heavy color-octet vector
particle $G^{'}$~\cite{ttH_theory}. Considering this scenario, we can exclude a region in
the $[m_{t^{'}}, M_{H}]$ parameter space, excluding for example values below a
Higgs boson mass of 146~GeV at a $t^{ '}$ mass of 355~GeV.

\section{\boldmath Searches for new physics in top quark decays}
We study the possibility of the top quark decaying into any
other quark but a $b$-quark by measuring the ratio of branching
fractions, and consider the possibility of the top quark decaying into another boson than the $W$ boson by performing a  search  for charged Higgs bosons.

In the SM, the top quark decays with a probability of almost 100\% into a $W$-boson and a $b$-quark. If, for example,  a fourth generation of quarks would exist, it could modify $B(t\rightarrow Wb)$ to a value below one. We study such a possibility, by measuring the ratio of branching fractions, defined as  
$R=\frac{B(t\rightarrow  Wb)}{B(t\rightarrow Wq)}$ with $q=b,s,d$. By analysing the
distribution of events with $0$, $1$ or $\ge2$ identified $b$-jets, a
simultaneous measurement of $\sigma_{t\bar{t}}$ and $R$ yields
$\sigma_{t\bar{t}}=8.18_{-0.84}^{+0.90}~{\rm (stat+syst)}\pm0.5~{\rm
  (lumi)}$~pb and $R=0.97_{-0.08}^{+0.09}$~\cite{Rb}.  We use this
result to set lower limits on $|V_{tb}|$ assuming unitarity of the
$3\times3$ CKM matrix, yielding $|V_{tb}|>0.89$ at 95\% C.L. Without
this assumption we can set an upper limit of $0.27$ on $(|V_{ts}|^2 +
|V_{td}|^2)/|V_{tb}|^2$ at 95\% C.L. The measured values of $R$ and $\sigma_{t\bar{t}}$ are fully consistent with the SM expectation. 

Not only the quark-part of the top decay can be modified by new physics, but also the bosonic part can be different. 
If, for example,  a charged Higgs boson exists with a mass smaller than the top quark
mass, the decays $t\rightarrow Wb$ and $t\rightarrow H^{+}b$ can
compete. Due to the different decay modes of $W$~boson and $H^{\pm}$ boson
the occurance of a charged Higgs boson leads to a different distribution of
events between various final states than in the SM. By comparing the
distribution of events in the lepton+jets, dilepton and $\tau$+lepton
final states, we are sensitive to $B(t\rightarrow
H^{+}b)$~\cite{xseccombi, Hplus_global}. 
We consider two
decay modes of charged Higgs bosons: $H^{+}\rightarrow \tau \nu$ and
$H^{+}\rightarrow c\bar{s}$, and the mixture of both decays. For the tauonic decaying charged Higgs,
we expect a decreasing number of events (disappearance) in the lepton+jets and dilepton final state and
a increasing number of  events (appearance) in the $\tau$+lepton channel for increasing $B(t\rightarrow H^{+}b)$. In case
of hadronic decaying charged Higgs bosons, all considered channels are
disappearance channels. We use two approaches to explore the
distribution of events between the final states. One is using
the ratio of the $t\bar{t}$ cross section measured in two exclusive final
states~\cite{xseccombi}. This has the advantage that many systematic
uncertainties, as for example on the luminosity, cancel. The second
approach uses the full information of all final states by performing a
global fit~\cite{Hplus_global}. Extra sensitivity can be gained here
by performing a simultaneous fit of $\sigma_{t\bar{t}}$ and
$B(t\rightarrow H^{+}b)$ for  tauonic decaying charged Higgs
bosons. Using this approach, we can set upper limits on
$B(t\rightarrow H^{+}b)$ between 0.13 for low charged Higgs masses and
0.26 for high masses at 95\% C.L. In case of a the hadronic decaying charged Higgs,
we set upper limits of 0.22 at 95\% C.L.

Besides the distribution of events between different final states, we also
study topological differences between events with tauonic
decaying charged Higgs bosons and SM $t\bar{t}$
events in the lepton$+$jets final
state~\cite{Hplus_topo}. Especially for high charged Higgs masses, the
upper limits on $B(t\rightarrow H^{+}b)$ are comparable with the global
fit method.

The limits we derived on $B(t\rightarrow H^{+}b)$ can be used to
set exclusion regions in the Minimal Supersymmetric Standard Model
(MSSM) parameter space $[M_{H^{+}},\tan \beta]$.  We consider three
different benchmark models, two of them are CP-conserving and one is
CP-violating. 
The CP-conserving models under study are the so-called MhMax and
no-mixing scenarios. 
By setting the stop-mixing parameter to be large (MhMax) or to zero
(no-mixing) a large parameter space in direction of the mass of the lightest Higgs
boson of the MSSM or a relatively restricted MSSM parameter space are
provided. 

\begin{figure}
\includegraphics[width=0.5\textwidth,clip=]{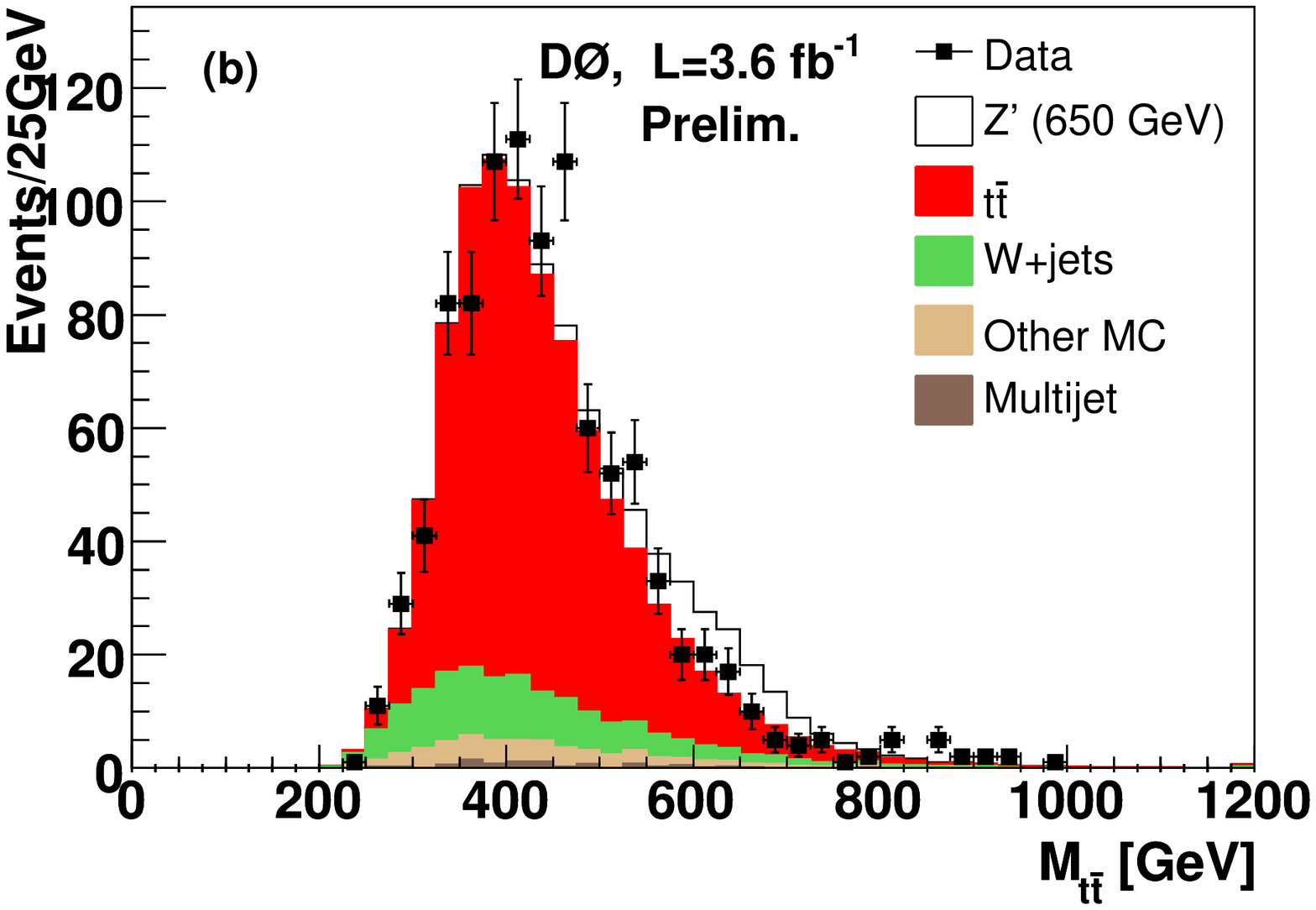}
\includegraphics[width=0.5\textwidth,clip=]{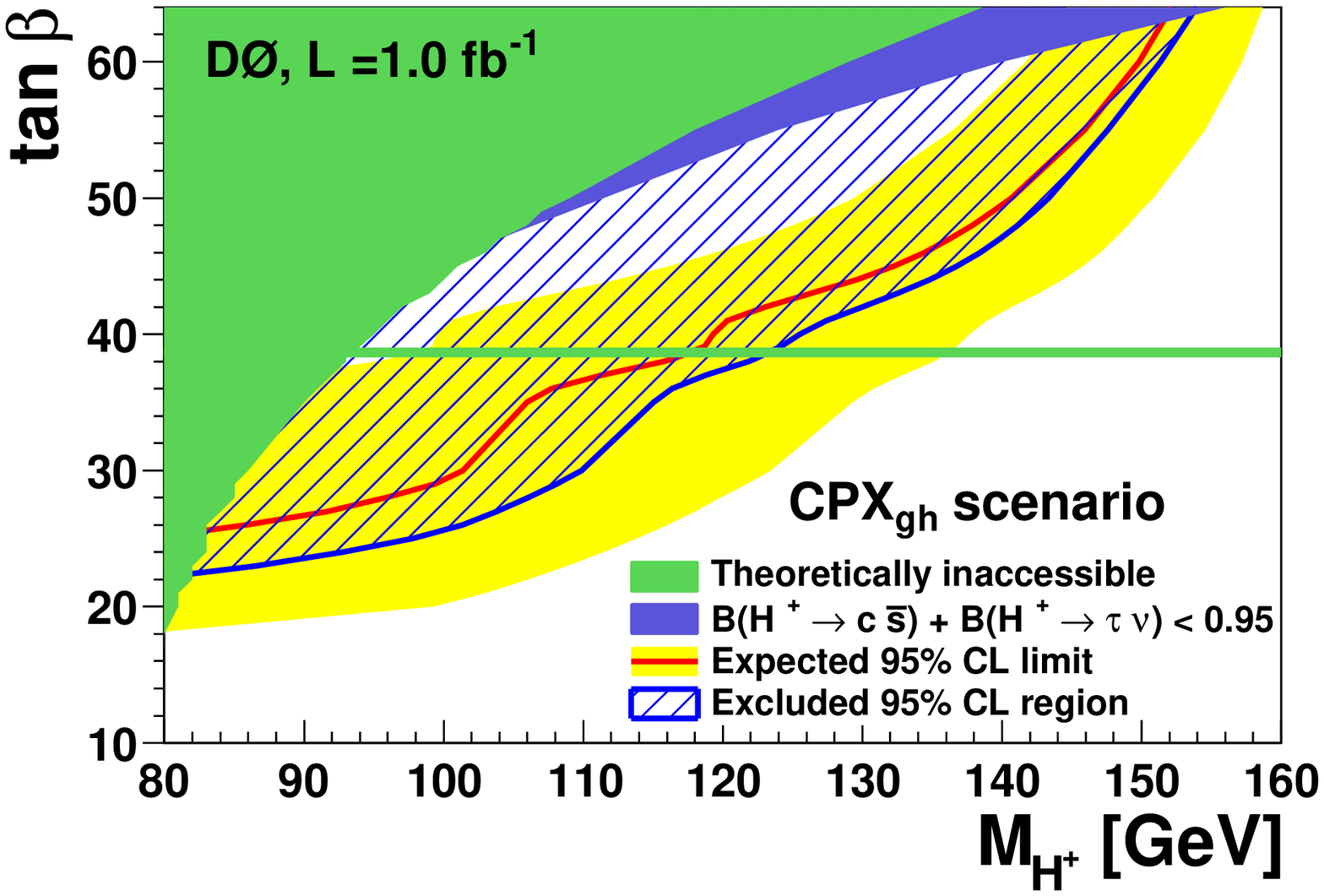}
\caption{\label{CPX_exclusion} Left: Invariant $t\bar{t}$ mass~\cite{reso_d0}. Right:
  Excluded region in $\tan \beta$ and $M_{H^{+}}$ in the MSSM
  for the strangephilic CPX model~\cite{strangephilicCPX,Hplus_global}.}
\end{figure}

As a CP-violating model we consider a strangephilic CPX
scenario~\cite{strangephilicCPX}, where the charged Higgs boson decay
into charm and strange quarks is enhanced by introducing a hierarchy
between the first two and the third generation of sfermions. In this
model, we can exclude charged Higgs masses up to 154~GeV for large
$\tan \beta$. Figure~\ref{CPX_exclusion} shows the exclusion region in
this model.

\section{\boldmath Searches for new physics in single top quark production}
Besides doing searches in the $t\bar{t}$ production channel, it is
also interesting to search for deviations from the SM expectation in
single top production. At the Tevatron,  single top quarks are produced
dominantly via
the electroweak interaction in s- and t-channel diagrams in the SM. Two examples of searches in the single top sector are presented here, namely the
search for a heavy charged Higgs bosons and a heavy gauge boson
$W^{'}$.

If a charged Higgs boson heavier than the top quark would exist, the
process $p\bar{p}\rightarrow H^{+} \rightarrow t\bar{b}$ could take
place. Besides leading to an increased number of expected events than in the SM
prediction only,  such a process would lead to a bump in the
invariant mass distribution of the $t \bar{b}$ final state. At D0, a search for
heavy charged Higgs bosons with masses between $180$ and $300$~GeV has
been performed using $0.9~{\rm  fb}^{-1}$ of integrated
luminosity~\cite{Hplus_heavy}. To allow for a full reconstruction of
the final state, only events with exactly two jets are
considered for this search. No evidence for a heavy charged Higgs boson is
found. Upper limits on $\sigma(p\bar{p}\rightarrow
H^{+})\times B(H^{+}\rightarrow t\bar{b})$ are set using three different
types of two-Higgs-doublet models (2HDM). For type I 2HDM models, where only
one of the doublets couples to fermions, a region in the $[M_{H^{+}},
\tan \beta]$ plane can be excluded. 

 Similar to the heavy charged Higgs boson search, the invariant mass
 distribution of the $t \bar{b}$ final states can also be used to
 search for heavy gauge bosons $W^{ '}$. We performed a search for
 $W^{'}$ bosons with $0.9~{\rm fb}^{-1}$ of D0 data~\cite{Wprime}. Two different
 scenarios have been considered: One where the $W^{'}$ has a
 left-handed coupling, leading to interferences with the
 SM $W$ boson, and one where the $W^{'}$ has a right-handed
 coupling. We can exclude $M_{W^{'}}<731$~GeV for $W^{'}$ with left
 handed couplings. In case of $W^{'}$ with right-handed couplings, we
 exclude   $M_{W^{'}}<739$~GeV if the $W^{'}$ decays to leptons and
 quarks and $M_{W^{'}}<768$~GeV if the $W^{'}$ decays only into quarks.

\section{Conclusion and Outlook}
In this report, a representative extract of searches for physics
beyond the SM using top quark events has been presented. We used up to $3.6~{\rm fb}^{-1}$ of data
collected with the D0 detector, yielding high sensitivity for various
new models. Until today, more than $6~{\rm fb}^{-1}$ of data have been
collected. With improving techniques and the increasing statistics the
top quark will remain an ideal particle to look for new physics.

\end{document}